\documentclass[prl,twocolumn,showpacs]{revtex4}

\usepackage{amssymb,amsmath,bm,epsfig,graphics,subfigure}

\begin{document}

\title{Commensuration and Interlayer Coherence in Twisted Bilayer Graphene}
\author{E. J. Mele}
    \email{mele@physics.upenn.edu}
    \affiliation{Department of Physics and Astronomy \\ University of Pennsylvania, Philadelphia PA 19104}
\date{\today}

\begin{abstract}
The low energy electronic spectra of rotationally faulted graphene bilayers are studied using a long wavelength theory
applicable to general commensurate fault angles. Lattice commensuration requires low energy electronic coherence across a fault
and preempts massless Dirac behavior near the neutrality point. Sublattice exchange symmetry distinguishes two families of
commensurate faults that have distinct low energy spectra which can be interpreted as energy-renormalized forms of the spectra
for the limiting Bernal and AA stacked structures. Sublattice-symmetric faults are generically fully gapped systems due to a
pseudospin-orbit coupling appearing in their effective low energy Hamiltonians.
\end{abstract}

\maketitle

    Coherent interlayer electronic motion in multilayer graphenes play a crucial role in their low energy properties \cite{McClure}.
This physics is well understood for stacked structures with neighboring crystallographic axes rotated by multiples of $\pi/3$,
including AB (Bernal), AA, ABC stackings and their related polymorphs \cite{MM }. Here the interlayer coupling scale typically
exceeds $0.5 \, {\rm eV}$ and preempts the massless Dirac physics of an isolated graphene sheet. Indeed experimental work on
Bernal stacked bilayers \cite{McCann ,Ohta ,Castro ,Li } demonstrates that their electronic properties
 are radically different from those of a single layer \cite{Novoselov ,Zhang }. Yet, recent experimental work has revealed a
family of multilayer graphenes that show only weak (if any) effects of their interlayer interaction. These include graphenes
grown epitaxially on the SiC $(000 \bar{1})$ surface \cite{Berger ,Hass ,Hass2}, mechanically exfoliated folded graphene
bilayers \cite{Schmidt } and graphene flakes deposited on graphite \cite{LiAndrei }. A common structural attribute of these
systems is the rotational misorientation of their neighboring layers at angles $\theta \neq n \pi/3$. A continuum theoretic
model has suggested that misorientation by an arbitrary fault angle induces a momentum mismatch between the tips of the Dirac
cones in neighboring layers suppressing coherent interlayer motion at low energy \cite{LpD }. In this interpretation, the Dirac
points of neighboring layers remain quantum mechanically {\it decoupled} across a rotational fault \cite{Hass2,LpD ,Latil
,Campanera ,Shallcross ,GTdL }  accessing two dimensional physics in a family of three dimensional materials.

This Letter presents a long wavelength theory of electronic motion in graphene bilayers containing rotational faults at
arbitrary commensurate angles. I find that the Dirac nodes of these structures are directly coupled across any commensurate
rotational fault, producing unexpectedly rich physics near their charge neutrality points. The theory generalizes previous
approximate analyses \cite{LpD } by treating the lateral modulation of the interlayer coupling between rotated layers which is
essential for understanding the low energy physics. Importantly, commensurate rotational faults occur in two distinct forms
distinguished by their sublattice parity. Structures that are even under sublattice exchange (SE) are generically gapped
(nonconducting) materials while those that break SE symmetry have two massive (curved) bands contacted at discrete Fermi
points. Both these behaviors derive from the the spectral properties of AA and Bernal stacked structures, and can be understood
as energy-renormalized versions of these limiting cases. The gap in the faulted sublattice-symmetric states appears as a new
feature specific to the faulted structures due to a pseudospin-dependence of the transmission amplitude across a twisted
bilayer. These results provide the appropriate low energy Hamiltonian(s) for rotationally faulted bilayers superseding the
massless Dirac model of an isolated sheet.

The crystal structure of two dimensional graphene (Fig. 1) has a Bravais lattice spanned by two primitive translations
$t_1=e^{-i \pi /6}$ and $t_2=e^{i \pi /6}$ with sublattice sites at $\tau_{A(B)} = 0(1/\sqrt{3})$. We consider rotational
stacking faults that fix overlapping A-sublattice sites at the origin and rotate one layer through angle $\theta$ with respect
to the other, with translation vectors $(t'_1, t'_2) = e^{i \theta}(t_1,t_2)$ and  basis $\tau'_{A(B)} = e^{i \theta}
\tau_{A(B)}$. A commensurate rotation occurs when $T_{m,n} = mt_1 + nt_2 = m' t'_1 + n' t'_2 = T'_{m',n'}$ i.e. at discrete
angles indexed by two integers $m$ and $n$ where $\theta(m,n) = \arg[(m e^{-i \pi/6} + n e^{i \pi/6})/(n e^{-i \pi/6} + m e^{i
\pi/6})]$. In this notation AA stacking (all sites in neighboring layers eclipsed) has $\theta =0$ and Bernal stacking has
$\theta = \pi/3$. Small angular deviations from the Bernal structure have indices $m=1$ and large $n$. The $\sqrt{13} \times
\sqrt{13}$ structures with $\theta = 30^\circ \pm 2.204^\circ$ structures observed by electron diffraction from epitaxial
graphene on the Si $(000\bar{1})$ face correspond to $(m,n) = (1,3)$ and $(m,n) = (2,5)$\cite{nonprimitive}.

\begin{figure}
\begin{center}
  \includegraphics[angle=0,width=\columnwidth]{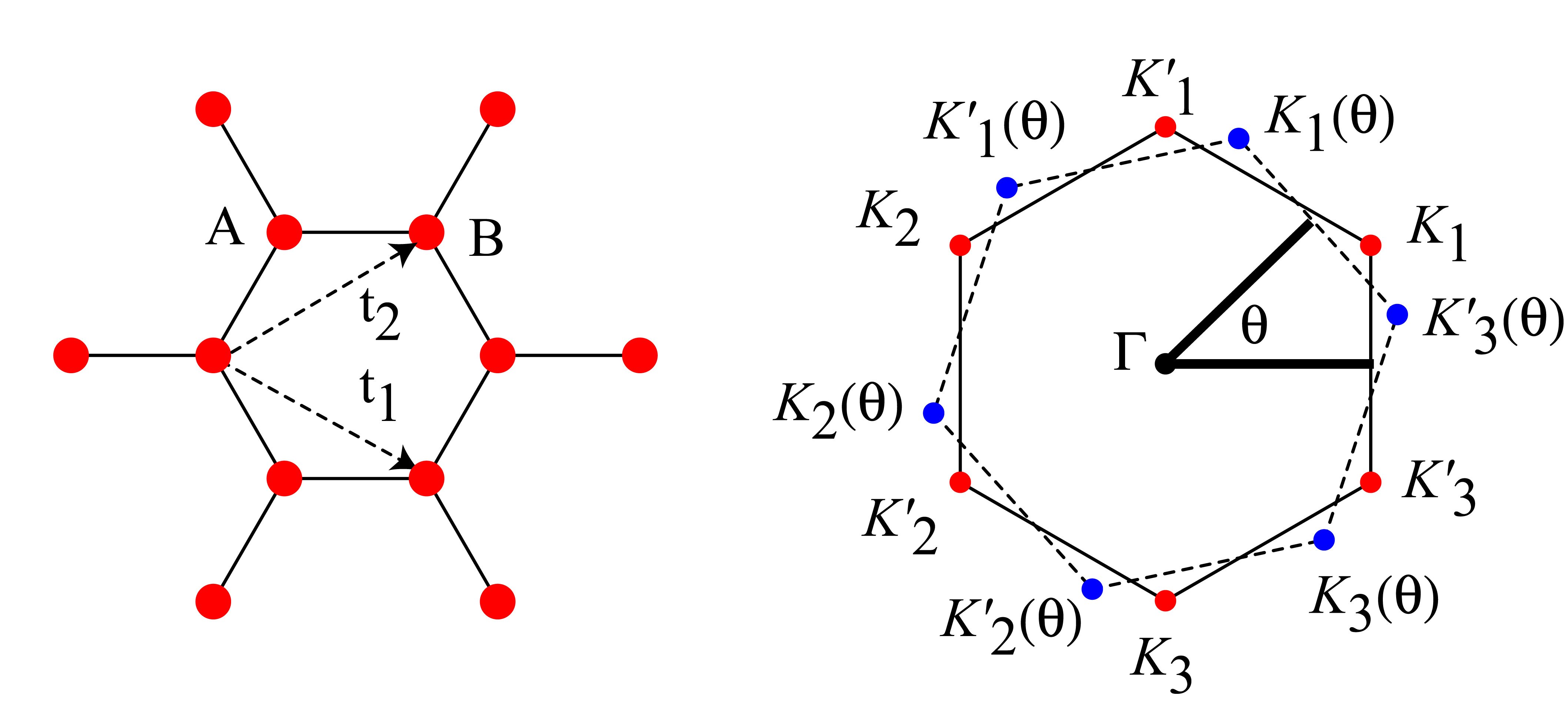}
  \caption{\label{lattice} (left) Lattice structure of graphene with two sites in the primitive cell ($A$ and $B$) and  primitive translations
 $t_1$ and $t_2$.  (Right) Brillouin zones for the two layers in a rotational fault: the  Brillouin zone corners labelled $K_m$
  and $K'_m$  are rotated by angle $\theta$ to the points $K_m(\theta)$ and $K_m'(\theta)$ in the neighboring layer.}
\end{center}
\end{figure}

Commensurate faults occur in two families determined by their sublattice exchange (SE) symmetry. With the A-sublattice sites at
the origin, a commensuration is SE symmetric if B-sublattice sites are coincident at some other lattice position in the
primitive cell. This occurs when $\tau_B + T_{p,q} = \tau_{B'} + T'_{p',q'}$ for integers $(p,q)$ and $(p',q')$, requiring
integer solutions to $p = (m - n + 3mq)/(3n)$.  This occurs only when $m-n$ is divisible by $3$ and then the coincident $B(B')$
sublattice site occurs at one of three possible threefold-symmetric Wyckoff positions of the cell (e.g. $T_{m,n}/3$ in Fig. 2).
The remaining Wyckoff positions are occupied by the A(A')-sublattice sites (the origin) and by overlapping hexagon centers
(H,H'). When $m-n$ is not divisible by 3 the only coincident site is the A-site at the origin, and the remaining two threefold
symmetric Wyckoff positions are occupied by B-sublattice atoms of one layer aligned with the hexagon centers (H') of its
neighbor.  Rotational faults at angles  $\bar \theta = \pi/3 - \theta$ form commensuration partners with primitive cells of
equal areas but opposite sublattice parities. Fig. 2 illustrates this situation for two partner commensurations at $30^\circ
-8.213^\circ$ ($(m,n)=(1,2)$) (left) and $30^\circ + 8.213^\circ$ ($(m,n)=(1,4)$) (right). The limiting cases of Bernal (odd)
and AA (even) stackings form the shortest period commensuration pair.

\begin{figure}
\begin{center}
  \includegraphics[angle=0,width=\columnwidth]{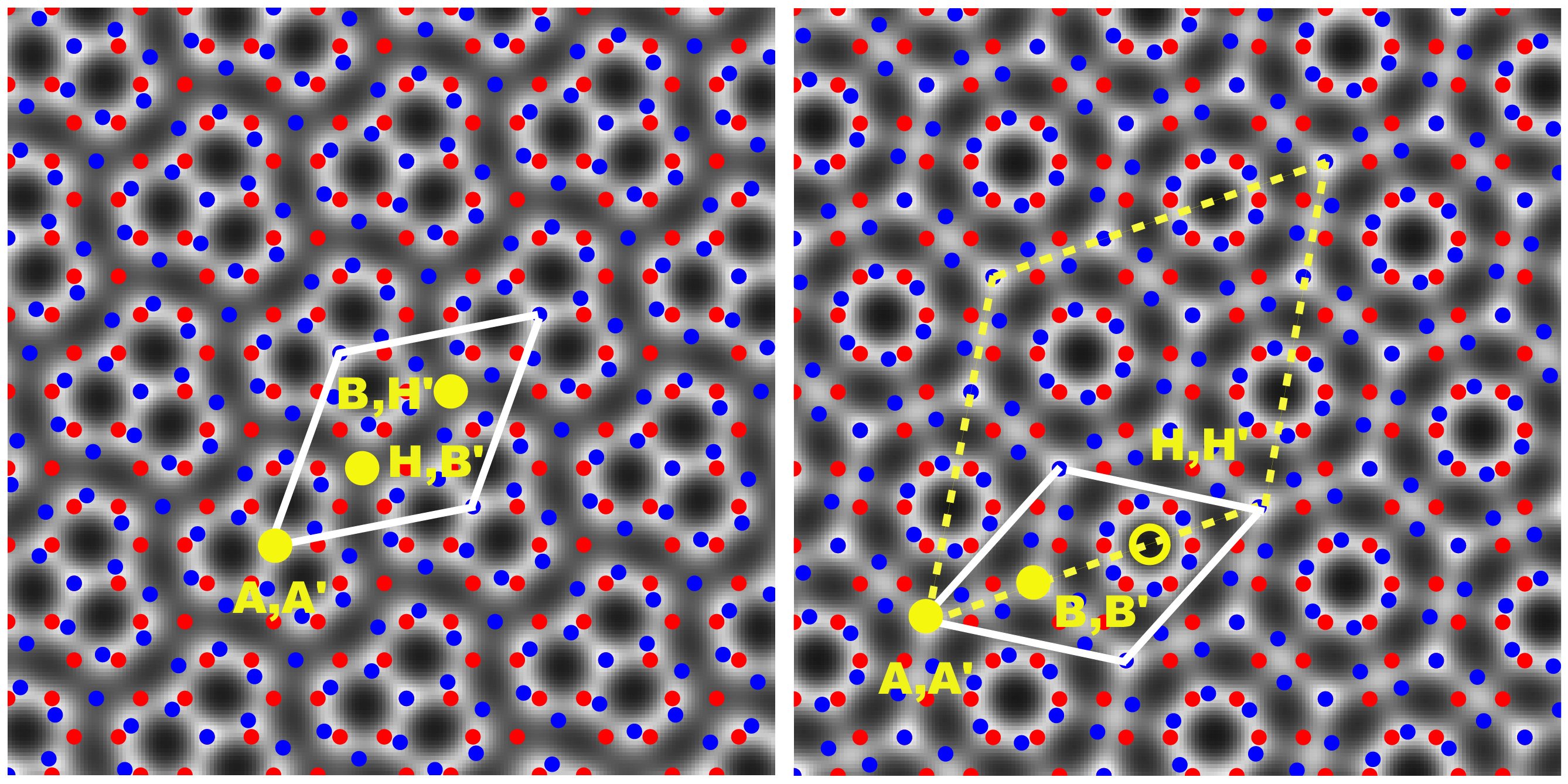}
  \caption{\label{hoplattice} Geometry of the commensuration cells for bilayers faulted at $21.787^\circ$(left) and
  $38.213^\circ$ (right). Red and blue dots denote the atom positions in the two layers. The white rhombuses denote
  primitive commensuration cells with the same area for these two structures. The dashed yellow rhombus denotes
  a $\sqrt{3} \times \sqrt{3}$ nonprimitive cell. The left hand structure is SE odd,
  with coincident atomic sites only on the A(A')-sublattice
   at the origin, the right hand structure is SE even with coincident sites on the A(A') and B(B') sublattices at threefold symmetric
   Wyckoff positions in the primitive cell and overlapping hexagon centers (H,H'). The
   density plot give the magnitude of the interlayer hopping potential discussed in the text.}
\end{center}
\end{figure}

 Because of the rotation, the Brillouin zones of the two layers have different orientations (Fig. 1(b)) shifting their zone corners
 ($K_m$, $K'_m$) to rotated counterparts ($K_m(\theta)$, $K'_m(\theta)$). The low energy
electronic states of the decoupled layers have isotropic conical dispersions near each of these points with $E(q) = \pm \hbar
v_F |q|$ where $q$ is the crystal momentum measured relative to the corner and $v_F$ is the Fermi velocity. These spectra are
described by a pair of massless Dirac Hamiltonians for the $K$ and $K'$ points of the two layers \cite{DPD }. Interlayer
coupling is studied using a long wavelength theory that represents the low energy states as spatially modulated versions of the
orthogonal zone corner Bloch states of the two layers, i.e. $\Psi(\vec r) = \sum_{\alpha} \psi_{K,\alpha} (\vec r) u_\alpha
(\vec r) $. In the first star approximation, appropriate to the interlayer coupling problem, the basis states are the Bloch
waves $\psi_{K,\alpha} = (1/\sqrt{3}) \sum_m e^{i \vec K_m \cdot (\vec r - \vec \tau_\alpha)}$ retaining reciprocal lattice
vectors that constrain the sum in $\psi_{K,\alpha}$ to the three equivalent corners of the Brillouin zone. The coupling between
layers is derived from an interaction functional of the form
\begin{eqnarray}
U = (1/2) \int \, d^2r \, T_{\ell}(\vec r) |\Psi_1(\vec r) - \Psi_2(\vec r)|^2
\end{eqnarray}
which correlates the amplitudes and phases of the Bloch waves in neighboring layers. Here $T_{\ell}$ is a periodic coupling
potential with the translational symmetry of the commensuration cell. Using Eqn. (1) one finds that the interlayer interaction
energy can be expressed in terms of the Fourier transforms of the slowly varying fields $u_{\alpha}$
\begin{eqnarray}
 \frac{U_{\rm int}}{N} = \frac{{\cal A}_o}{24 \pi^2}   \int d^2 q \, \sum_{\alpha, \beta} \sum_{m ; m' } \,
    \,  e^{i \vec K_m \cdot \vec \tau_\alpha}
 e^{-i \vec K_{m'} \cdot \vec \tau'_\beta}   \nonumber\\
       \times \sum_{\vec{\cal G}} \left( t(\vec{\cal G})  u^*_{1,\alpha}(\vec q) u_{2,\beta}(\vec q + \vec K_m - \vec K_{m'} - \vec{\cal G})
         + {\rm c.c.} \right)
 \end{eqnarray}
 where $N$ is the system size, ${\cal A}_o$ is the area of a graphene primitive cell and
 $t(\vec {\cal G})$ is the Fourier transform of the interlayer potential $T_{\ell}(\vec r)$ on the reciprocal lattice
 of the commensuration cell $\vec {\cal G}$.

The continuum theory of reference \cite{LpD } is recovered from Eqn.2 by retaining only its $\vec {\cal G}=0$ terms, thus
treating the interlayer coupling as {\it spatially uniform}.  In this approximation the states near the tip of the Dirac cone
in one layer are coupled to three pairs of states at energies $\pm W^* = \pm \hbar v_F |K_m - K_m(\theta)|$ in its neighbor. At
low energies, the effect of this coupling can be treated perturbatively, preserving the Dirac nodes of two isotropic
velocity-renormalized layer-decoupled Dirac Hamiltonians.

However, new physics arises from the $\vec {\cal G} \neq 0$ contributions in Eqn. 2 which mediate a direct coupling between the
Dirac nodes of neighboring layers and prevent massless low energy behavior. To study it, note that the reciprocal lattice of
the bilayer is spanned by momenta with {\it four} integer indices $\vec {\cal G} = p \vec G_1 + q \vec G_2 + p' \vec G'_1 + q'
\vec G'_2$ \cite{reciprocal }. Momentum conserving couplings between $K$ points in neighboring layers occur when $K_m -
K_{m'}(\theta) = \vec {\cal G}(p,q,p',q')$ with the angle $\theta(m,n)$ specified. This is an {\it interlayer umklapp} process
where the spatial modulation of $T_{\ell} (\vec r)$ provides precisely the transverse momentum required to transport an
electron between the Dirac nodes of neighboring layers. The momentum matching condition requires integer $p$ solutions to $ p =
(m-n)/3n + qm/n$, and occurs only for {\it supercommensurate} structures with nonzero ${\rm mod}(m,3)={\rm mod}(n,3)$.
Importantly if this condition is not satisfied, momentum-conserving interlayer couplings still occur, but instead through the
analogous {\it intervalley} umklapp process, i.e. $K'_m - K_{m'}(\theta) = \vec {\cal G}(p,q,p',q')$. These two possibilities
are complementary and mutually exclusive: one or the other must occur if the rotational fault is commensurate. These two
criteria distinguish SE-even and SE-odd structures, so that the SE-even structures require direct $K-K(\theta)$ coupling and
SE-odd structures $K-K'(\theta)$ coupling.

To understand the consequences of the interlayer interaction one requires a theory for the Fourier coefficients $t(\vec {\cal
G})$ in Eqn. (2). These can be calculated from atomistic models, but their relevant properties are determined by symmetry. Note
that the coupling function $T_{\ell}(\vec r)$ is a real periodic function with the translational symmetry of the commensuration
cell. The structure function for the $\mu$-th layer,  $n_\mu(\vec r) = \sum_{m \in [1]} \sum_{\alpha} e^{i \vec G_{\mu,m} \cdot
(\vec r - \vec \tau_{\mu,\alpha})}$ superposes the six plane waves of the lowest star of reciprocal lattice vectors $\vec
G_{\mu,m}$ producing a standing wave with maxima on atom sites and minima in hexagon centers. A useful model for the interlayer
coupling potential is $T_{\ell} (\vec r) = C_o \exp[C_1 n(\vec r)]$  where $n(\vec r) =n_1 +n_2$ and $C_0$ and $C_1$ are
constants; $T_\ell$ is a superlattice-periodic function with maxima for coincident sites and with exponential suppression in
regions that are out of interlayer registry. The grayscale plot in Fig. 2 show the spatial distribution of $T_{\ell}(\vec r)$
where $C_1$ is determined by matching the decay of the hopping amplitude between neighboring layer atoms as a function of small
lateral offsets. This density plot shows that the interlayer amplitudes between rotated layers have coherent structures in the
forms of fivefold rings (from overlapping misaligned hexagons) that are arranged to form two dimensional space-filling
patterns. SE-odd structures are symmetric under threefold rotations while the SE-even structures retain  a sixfold symmetry.
The separable form $T_\ell(\vec r) = f_1(\vec r) f_2(\vec r)$ allows one to deduce a scaling rule for the Fourier coefficients:
$t(\vec {\cal G}) \approx (ae^{-b/N_c}/N_c) \sum_{\mu \in [1]} f_2(\vec r_\mu) e^{-i \vec{\cal G} \cdot \vec r_\mu}$ where the
sum is over atomic sites in layer 1, $a$ and $b$ are constants, and $N_c$ is the number of graphene cells (per layer) in the
commensuration cell. For large $N_c$ the prefactor decays as a power law of the cell size reflecting the fraction of atomic
sites in good interlayer registry while the sum decays quickly as a function of $N_c$ because of cancelling phases in its
argument.

The interlayer Hamiltonian can be expressed by a $3 \times 3$ array of scattering amplitudes derived from the $t(\vec {\cal
G})$'s giving the allowed transitions $K_m \rightarrow K_{m'}(\theta)$. Three fold rotational symmetry requires that this
matrix has the form
\begin{eqnarray}
\hat {\cal V}_{\rm ps} = \left(\begin{array}{ccc}
  V_0 & V_1 & V_2\\
  V_2 & V_0 & V_1 \\
  V_1 & V_2 & V_0 \\
\end{array} \right) \nonumber
\end{eqnarray}
where the pseudopotential coefficients $V_i$ are are matrix elements of $T_\ell$. Completing the sum in Eqn. 2 projects this
into the sublattice (pseudospin) basis and gives the $2 \times 2$ interlayer transition matrices $\hat {\cal H}_{\rm int}$ seen
by the Dirac fermions. The low energy Hamiltonian for an SE-even bilayer can be expressed as a $4 \times 4$ matrix (acting on
the two sublattice and two layer degrees of freedom)
\begin{eqnarray}
\hat {\cal H}_{\rm even} = \left( \begin{array}{cc}
  -i \hbar \tilde{v}_F \sigma_1 \cdot \nabla
 & \hat {\cal H}^+_{\rm int} \\
  (\hat {\cal H}^+_{\rm int})^\dagger & -i \hbar \tilde{v}_F \sigma_2 \cdot \nabla
 \\
\end{array} \right)
\end{eqnarray}
and for the SE-odd bilayer
\begin{eqnarray}
\hat {\cal H}_{\rm odd} = \left( \begin{array}{cc}
  -i \hbar \tilde{v}_F \sigma_1 \cdot \nabla
 & \hat {\cal H}^-_{\rm int} \\
  (\hat {\cal H}^-_{\rm int})^\dagger & i \hbar \tilde{v}_F \sigma_2^* \cdot \nabla
 \\
\end{array} \right)
\end{eqnarray}
where $\sigma_n$ are Pauli matrices acting in the sublattice pseudospin basis of the $n-th$ layer and $\tilde{v}_F$ is the
renormalized Fermi velocity. The interlayer matrices $\hat {\cal H}^\pm_{\rm int}$ are
\begin{eqnarray}
\hat {\cal H}^+_{\rm int} = {\cal V} e^{i \vartheta} \left(\begin{array}{cc}
  e^{i \varphi/2} & 0 \\
  0 & e^{- i \varphi/2} \\
\end{array} \right) ,\,\,\, \hat {\cal H}^-_{\rm int} = {\cal V} e^{i \vartheta} \left(\begin{array}{cc}
  1 & 0 \\
  0 & 0 \\
\end{array} \right)
\end{eqnarray}
$\hat {\cal H}^+_{\rm int}$ shows that interlayer motion of an electron for SE-even faults requires a unitary transformation of
its $(A,B)$ sublattice amplitudes  represented as an {\it xy rotation of its pseudospin through angle $\varphi$}. This angle is
{\it not} defined geometrically by the fault angle $\theta$ but rather is determined by the relative magnitudes of the three
pseudopotential matrix elements $V_i$.  By contrast interlayer motion across a sublattice asymmetric fault involves only the
amplitudes on its dominant (eclipsed) sublattice. The continuum model of \cite{LpD } is recovered by setting $\hat {\cal
H}_{\rm int}=0$.

In either case, below an energy scale ${\cal V}$ the electronic spectra deviate from the massless Dirac form and inherit
curvature from the interlayer coupling as shown in Fig. 3. ${\cal V} \sim 10$ meV for commensurations at $\theta = 30^\circ \pm
8.213^\circ$ with $N_c =7$ graphene cells per layer in their commensuration cells. Nevertheless the forms of these spectra
apply generally to any pair of commensuration partners. SE odd faults mix the degenerate Dirac bands gapping one pair on the
scale ${\cal V}$, leaving a second pair of massive (curved) bands in contact at $E=0$.  By contrast, SE even structures are
fully gapped where the gap arises entirely from the pseudospin rotation in Eq. 5. Indeed for ${\varphi}=0$ these spectra
consist of a pair of Dirac cones offset in energy by a bonding-antibonding splitting, and intersecting at $E=0$ on a circle in
reciprocal space. The pseudospin rotation lowers the symmetry of the bilayer and produces an avoided crossing of these states.
For SE-even rotated bilayers the interlayer coupling describes a type of spin orbit coupling with the sublattice pseudospin
index playing the role of the spin.
\begin{figure}
\begin{center}
  \includegraphics*[angle=0,width=\columnwidth]{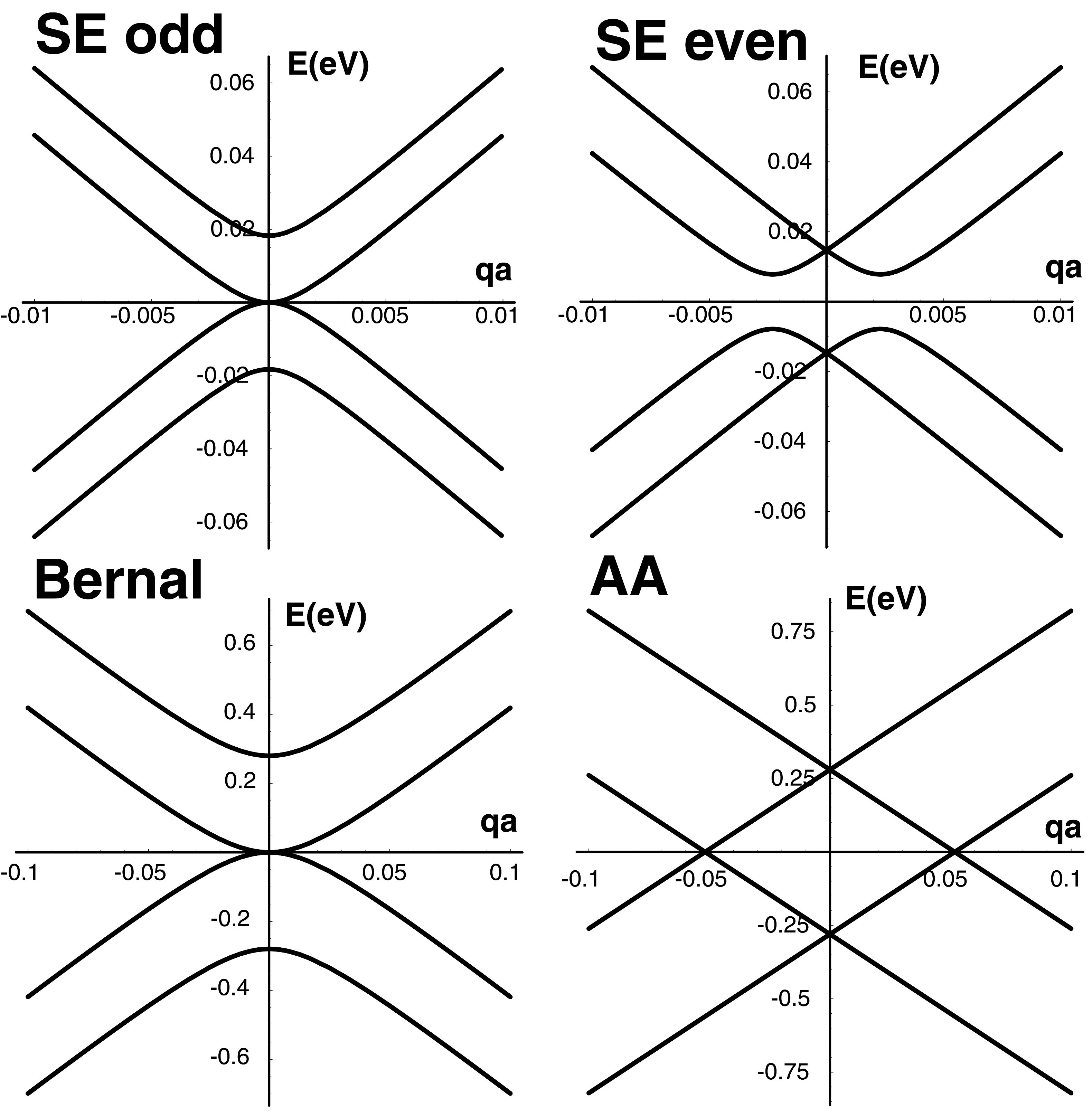}
  \caption{\label{spectra7} Low energy electronic spectra for SE-odd and SE-even faulted bilayers are illustrated
  using partner commensurations at rotation angles $\theta = 21.787^\circ$
  (odd, left) and $\theta = 38.213^\circ$ (even, right). These spectra are symmetric under rotations in momentum space. SE odd
  faults have massive bands that contact at Fermi points (left) and SE-even faults are gapped (right). The lower
  row gives the spectrum for a Bernal bilayer (left) and for an AA bilayer, which show related spectral properties.}
\end{center}
\end{figure}

These behaviors have precise analogs for the limiting cases of unfaulted Bernal and $AA$ stacked layers. The Bernal bilayer
(lower left Fig. 3) has exactly the structure found for SE-odd faults but on an inflated energy scale $\sim 0.5$ eV, reflecting
the full alignment of all atoms on a single sublattice. Similarly, for $AA$ bilayers (lower right Fig. 3) interacting Dirac
cones are displaced in energy  but {\it without} a pseudospin rotation, so they intersect on rings at $E=0$. All intermediate
commensurate site-centered rotational faults display either energy-renormalized Bernal-like or $AA$-like low energy spectra;
the reduction of the energy scale is a measure of the loss of interlayer registry in the faulted bilayer. This correspondence
can be deduced from the lattice symmetries of the density plots shown in Fig. 2.

Since ${\cal V} < W^*$ rotational faults open an energy ``window" ${\cal V} < E < W^*$ in which the physics is well described
by decoupled two-dimensional systems before their interlayer coherence is apparent.  The theory of small angular deviations
from Bernal stacking \cite{LpD } can be understood as a collapse of the energy scale ${\cal V}$ relative to $W^*$. Yet, physics
at the scale ${\cal V}$ is accessible to experimental probes and highly relevant to electrostatic gating and charge transport
in structures derived from multilayer graphenes \cite{Schmidt ,Castro }  particularly those with faults near $\theta = \pi/6$
found in epitaxial graphenes on SiC $(000\bar 1)$\cite{Hass2}. Additionally, the many body physics of these systems depends
crucially on the low energy structure of these spectra \cite{MBSM } and it can be expected to be quite different for different
stacking sequences. Thus one can regard faulted multilayer graphenes as presenting a family of new materials with properties
that interpolate between single layer graphene and bulk graphite in an understandable and hopefully controllable way.

I thank C.L. Kane, S. Kim and A.M. Rappe for their helpful comments. This work was supported by the Department of Energy under
grant DE-FG02-ER45118.


\begin{thebibliography}{11}
\bibitem{nonprimitive} This is indexed to a nonprimitive two dimensional cell. The related
{\it structural} primitive cell is obtained by using the deflation rule: $(m_p,n_p) = (m-n,m+2n)/3$ giving $(m_p,n_p) = (-1,4)$
which has the same area as the $(m,n) = (1,3)$ cell.
\bibitem{reciprocal } $\vec G_i$ and $\vec G'_j$ are primitive reciprocal lattice vectors of each sheet. One could also index $\vec
{\cal G}$ using two primitive reciprocal lattice vectors dual to the full commensuration cell, but the four index formulation
is more powerful.
\end{thebibliography}
\end{document}